\documentclass[preprint,showpacs,preprintnumbers,amsmath,amssymb]{revtex4}

\usepackage{graphicx}
\usepackage{bm}

\newcommand{\E}[1]{Eq.~(\ref{#1})}
\newcommand{\F}[1]{Fig.~\ref{fig:#1}}
\newcommand{\sle}{ \mathop{}_{\textstyle \sim}^{\textstyle <} }

\begin{document}

\preprint{nlin.CD/000000}

\title{Critical exponents of Nikolaevskii turbulence} 

\author{Dan Tanaka}
 \email{dan@ton.scphys.kyoto-u.ac.jp}
\affiliation{%
Department of Physics, Graduate School of Sciences, Kyoto University, Kyoto 606-8502, Japan
}%
\date{\today}%

\begin{abstract}   
We study the spatial power spectra of 
Nikolaevskii turbulence in one-dimensional space. 
First, we show that the energy distribution in wavenumber space 
is extensive in nature.  
Then, we demonstrate that, when varying a particular parameter,  
the spectrum becomes qualitatively indistinguishable from 
that of Kuramoto-Sivashinsky turbulence. 
Next, we derive the critical exponents of turbulent fluctuations. 
Finally, we argue that in some previous studies, 
parameter values for which this type of turbulence does not appear 
were mistakenly considered, and we resolve inconsistencies obtained in 
previous studies. 
\end{abstract}


\maketitle
The spontaneous formation of spatially periodic structure 
in reaction-diffusion systems 
was predicted by Turing in 1952 \cite{Tur} 
and experimentally confirmed many years later \cite{Cas, Ouy}. 
The so-called Turing mechanism is now widely accepted, and 
we can retrieve many papers by searching for the keyword `Turing pattern',  
including a large number written this century \cite{Kon}. 
Recently, we found evidence that 
the Turing instability in oscillatory systems 
can also cause an initially uniform state to evolve into a state 
characterized by spatiotemporal chaos 
instead of spatially periodic structure \cite{Dan03, Dan04}. 
This newly identified type of chemical turbulence is exhibited 
by the equation 
\begin{equation}
\partial_t \psi(x,t)=
-\partial_{x}^{2}[\epsilon-(1+\partial_{x}^{2})^{2}]\psi-(\partial
_{x}\psi)^2 \label{1},  
\end{equation}
which was derived from a class of oscillatory reaction-diffusion systems 
by means of a phase reduction technique \cite{Dan04}. 
An equivalent equation was proposed by V.~N.~Nikolaevskii 
as a model of seismic phenomena \cite{Nik}. 
The uniform steady state of \E{1}, $\psi=0$, is unstable with respect to 
finite-wavelength perturbations 
when the small parameter $\epsilon$ is positive. 
However, this instability does not lead to spatially periodic steady states, 
because the equation possesses a Goldstone mode, 
due to its invariance under transformations of the form 
$\psi \rightarrow \psi + {\rm const.}$, 
and the corresponding marginally stable long-wavelength modes 
interact with the unstable short-wavelength modes. 
As a consequence, spatially periodic steady states do not appear, and instead  
spatiotemporal chaos is realized supercritically \cite{Tri-Vel, Tri-Tsu}. 
Spatiotemporal chaos exhibiting a similar onset has been observed  
experimentally 
in electrohydrodynamic convection (``soft-mode turbulence'') 
in homeotropically aligned nematic liquidcrystals \cite{Kai} 
and 
numerically 
in Rayleigh-B$\acute{{\rm e}}$nard convection 
under free-free boundary conditions \cite{Xi}. 
In particular, \E{1} has been applied to the study of 
the former type of convective system.  
It is thus seen that this class of spatiotemporal chaos appears in many types of physical systems, and for this reason, studying \E{1} is important. 
In this paper, we study the statistical properties of the spatiotemporal chaos 
exhibited by \E{1} in one-dimensional space with periodic boundary conditions. 
Also, we argue that in some previous works on Nikolaevskii turbulence, 
values of $\epsilon$ that are in fact inappropriate for studying 
this type of turbulence were used. 

Equation~(\ref{1}) has two parameters, 
the bifurcation parameter $\epsilon$ and the system size $L$. 
First, we derive the $L$ dependence of the spatial power spectrum 
$S(q) \equiv  \langle |v_q|^2 \rangle $, where $v_q$ is the spatial Fourier transform of 
$v \equiv 2 \partial_x \psi$ and 
$ \langle  \rangle $ represents a long time average. 
The quantity $S(q)/L$ is plotted as a function of the wavenumber $q$ 
for $L=2^{9}, 2^{10}, 2^{11}$, and $2^{12}$ with $\epsilon=0.02$ 
in \F{extensive}. 
There it is seen that $S(q)/L$ possesses a universal form independent of $L$. 
This implies that the energy distribution in wavenumber space is 
an extensive quantity.  
In Ref.\cite{Xi-Ext}, the Lyapunov dimension and 
the Kolmogorov-Sinai entropy are studied 
for \E{1} in the cases $\epsilon=0.2, 0.5$, 
and it is shown that these too are extensive quantities. 
However, 
these values of $\epsilon$ are too large 
for \E{1} to exhibit the type of spatiotemporal chaos in which 
we are interested, as we show below. 
\begin{figure}
\resizebox{0.45\textwidth}{!}{\includegraphics{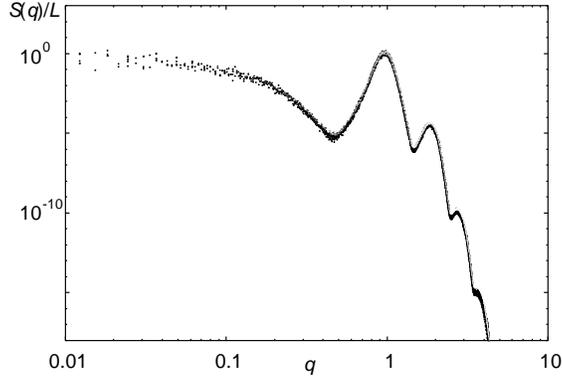}}
\caption{\label{fig:extensive}%
Spatial power density spectrum 
$S(q)/L$ as a function of the wavenumber $q$ 
for $L=2^{9}, 2^{10}, 2^{11}$, and $2^{12}$ with $\epsilon=0.02$. 
The fact that these plots fall on a universal curve independent of $L$ 
implies the extensive nature of the energy distribution in wavenumber space. 
}
\end{figure}

Second, we consider the $\epsilon$ dependence of the 
spatial power density spectrum $S(q)/L$.
In the following, we consider only the single system size $L=2^9$,  
because, as mentioned above, 
$S(q)/L$ is independent of $L$ when $L$ is sufficiently large. 
The peaks of the spectrum broaden and 
merge when $\epsilon$ increases, 
as shown in \F{ep-dep}. 
In particular, when $\epsilon$ is larger than about $0.1$, 
the spectrum is qualitatively indistinguishable from 
that of the Kuramoto-Sivashinsky (KS) equation, 
$\partial_t \psi(x,t)= 
-\partial_{x}^{2}(1+\partial_{x}^2)\psi-(\partial_{x}\psi)^2$ \cite{K84}. 
This can be understood as follows. 
The spatiotemporal chaos exhibited by \E{1} 
with a sufficiently small $\epsilon$ 
arises from the uniform steady state $\psi=0$, 
owing to the interaction between 
the weakly stable long-wavelength modes 
and the unstable short-wavelength modes.
The band of unstable modes has 
a width in wavenumber space of order $\epsilon^{1/2}$, 
lying on either side of $q=1$.
Therefore, unless $\epsilon^{1/2} \ll 1$,
the weakly stable and unstable bands of modes cannot be distinguished, 
and thus the situation is effectively the same as that for the KS equation, 
in which case chaos arises through 
interactions among unstable long-wavelength modes.
As a condition to insure that 
the unstable and weakly stable bands of \E{1} are sufficiently separated,  
we conjecture that $\epsilon^{1/2}$ must be 
at least one oreder of magnitude smaller than $1$. 
Hence, in order to clearly observe the characteristic 
Nikolaevskii chaos exhibited this equation, 
we believe that the $O(\epsilon) \leq 0.01$ is necessary. 
This leads us to conclude that the value $\epsilon=0.2$ 
and $\epsilon=0.5$ used in Ref.\cite{Xi-Ext} and Ref.\cite{Tor}
(which employs a wavelet decomposition) are too large to observe 
this type of spatiotemporal chaos 
and that the power spectrum found in those works is actually that of 
KS spatiotemporal chaos. 
In fact, it is shown below that the exponents of the $\epsilon$ scaling 
for the chaotic fluctuations of \E{1} do not converge 
for $O(\epsilon) \geq 0.1$. 
\begin{figure}
\resizebox{0.45\textwidth}{!}{\includegraphics{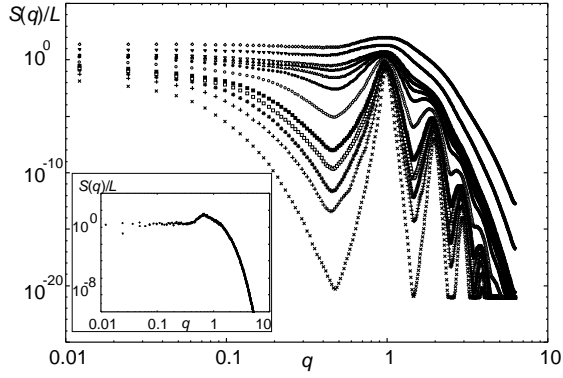}}
\caption{\label{fig:ep-dep}%
Spatial power density spectrum  
$S(q)/L$ as a function of the wavenumber $q$ 
for several values of $\epsilon$ with $L=2^9$. 
From top to bottom, we have 
$\epsilon=0.4, 0.2, 0.1, 0.08, 0.06, 0.04, 0.02, 0.01, 0.008, 0.006, 0.004$, 
and $0.002$.
The inset displays the spatial power density spectrum obtained from 
the well-known Kuramoto-Sivashinsky equation for the sake of comparison. 
}
\end{figure}

Noting that the spatiotemporal chaos exhibited by \E{1} results from 
the interaction between the long-wavelength modes near $q=0$
and the short-wavelength modes in the unstable band surrounding $q=1$, 
P.~C.~Matthews and S.~M.~Cox 
derived closed-form 
amplitude equations by hypothesizing that the behavior of the system 
can be described in terms of a quantity $v$ taking the form 
\begin{equation}
v = \epsilon^{3/4} A(X,T) e^{ix} + {\rm c.c.} + \epsilon f(X,T), \label{2}
\end{equation} 
where 
$A$ and $f$ represent the slowly varying amplitudes of the two sets of modes,
and we define 
$X \equiv \epsilon^{1/2} x$ and $T \equiv \epsilon^1 t$ \cite{Mat}. 
Also, H.~Fujisaka {\em et al.}
derived amplitude equations 
applicable in higher-dimensional spaces  
using the form \E{2} \cite{Fuj}. 
The validity of this form is supported by numerical results that 
show $\sqrt{ \langle v^2 \rangle } \propto \epsilon^{3/4}$ for $\epsilon \in [0.01,0.1]$ 
\cite{Mat}. 
However, 
based on an analysis of the time series of the spatial Fourier amplitude of 
turbulent fluctuations for \E{1} with the fixed value $\epsilon = 0.0001$,
Tribelsky and Tsuboi conjectured the scaling $\sqrt{\langle v^2 \rangle }
\propto \epsilon^{1/2}$ in Ref.\cite{Tri-Tsu}. 
Also, in Ref.\cite{Kli} it is shown that  
$\sqrt{ \langle v^2 \rangle } \propto \epsilon^{1}$ for $\epsilon \in [0.1,1]$. %
The discrepancy in the latter case seems to be easily accounted for, 
as it would appear that the result for the exponent reported in Ref.\cite{Kli}
had not yet converged, 
because the value of $\epsilon$ used there is too large. 
However, the situation is not so clear with regard to 
the apparent inconsistency reported in Ref.\cite{Tri-Tsu}, 
because the value of $\epsilon$ used there is certainly sufficiently small. 
Furthermore, we believe that 
the numerical results of Ref.\cite{Mat} 
are insufficient to 
establish the validity of the form given in \E{2} 
for the following reasons:  
1. The results were obtained for values of $\epsilon$ 
in a range of only one order, $\epsilon \in [0.01,0.1]$. 
It is quite likely that this small range is 
insufficient to yield a clear result for the power law exponent.  
2. Studying only the order parameter $\sqrt{ \langle v^2 \rangle }$, 
we are able to examine the validity of only the assumed exponent $3/4$ 
for the amplitude of $e^{ix}$. 
Verifying the validity of the other exponents 
requires a different approach. 
3. Employing the spatial coordinate $X=\epsilon^{1/2} x$ 
implies the assumption that the spatial scale of turbulent fluctuations 
is very much larger than that of the fundamental wave $e^{ix}$.  
This scale separation is ensured when $\epsilon^{1/2} \ll 1$. 
Taking this condition as implying that $\epsilon^{1/2}$ 
can be no greater than $0.1$, 
we obtain the requirement $O(\epsilon) \leq 0.01$
to guarantee sufficient separation of scales. 
Now, 
to resolve the inconsistency among the results of the previous studies and 
to examine the validity of \E{2},
we define some new order parameters and examine 
their $\epsilon$ dependence 
both for smaller values of $\epsilon$ and over a wider range of values of $\epsilon$ than in previous studies. 

First, as shown in \F{rms}, 
we find that the results for $\sqrt{ \langle v^2 \rangle }$ 
converge for $\epsilon \sle 0.1$, 
where we have $\sqrt{ \langle v^2 \rangle } \propto \epsilon^{3/4}$. 
These results indicate that if we wish to study the characteristic 
sptiotemporal chaos exhibited by \E{1}, we must choose a value of $\epsilon$ 
no greater than $0.1$. 
Then, as seen in \F{dq}, we find $\Delta q \propto \epsilon^{1/2}$, 
where $\Delta q$ is defined as the width at half maximum 
for the peak centered at $q \simeq 1$ of the spatial power density spectrum. 
This is evidence 
that the characteristic spatial scale of turbulent fluctuations 
is $\epsilon^{-0.5}$.
Finally, we present in Figs.~\ref{fig:sq0} and \ref{fig:sqc}
the $\epsilon$ dependence of the spatial power densities  
$S(q_0)/L$ and $S(q_c)/L$ for the two characteristic modes $q_0$ and $q_c$. 
These are the wavenumbers nearest to $q=0$ and $q=1$
(explicitly, $q_0=1 \times 2\pi/L$ and $q_c=81 \times 2\pi/L$) 
that can be realized in our system of size 
$L=2^9$ with periodic boundary conditions. 
From the figures it is seen that $S(q_0) \propto \epsilon^{3/2}$ and 
$S(q_c) \propto \epsilon^{1}$.
The former relation cannot be seen as clearly as the latter, 
because, for larger wavelength modes, 
fewer wavelengths are contained in the finite-size space, 
and therefore there is a larger statistical error in the result, 
which is obtained by integrating \E{1} over a finite interval of time. 
\begin{figure}
\resizebox{0.45\textwidth}{!}{\includegraphics{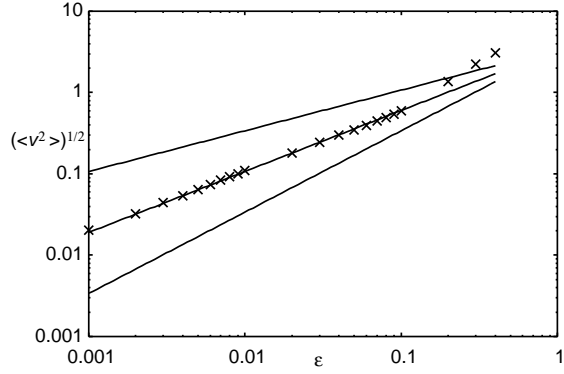}}
\caption{\label{fig:rms}%
$\epsilon$ dependence of $\sqrt{ \langle v^2 \rangle }$. 
The three lines, included for reference, have slopes of $2/4, 3/4$, and $4/4$.
}
\end{figure}
\begin{figure}
\resizebox{0.45\textwidth}{!}{\includegraphics{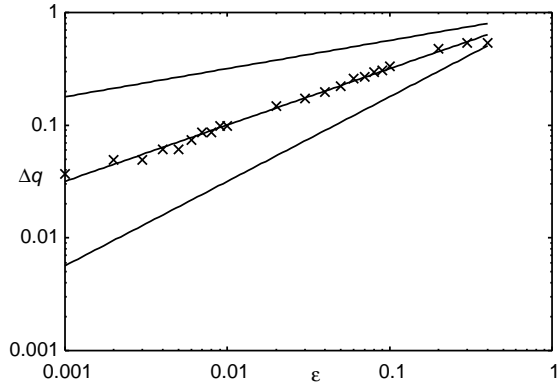}}
\caption{\label{fig:dq}%
$\epsilon$ dependence of $\Delta q$. 
The three lines, included for reference, have slopes of $1/4, 2/4$, and $3/4$.
}
\end{figure}
\begin{figure}
\resizebox{0.45\textwidth}{!}{\includegraphics{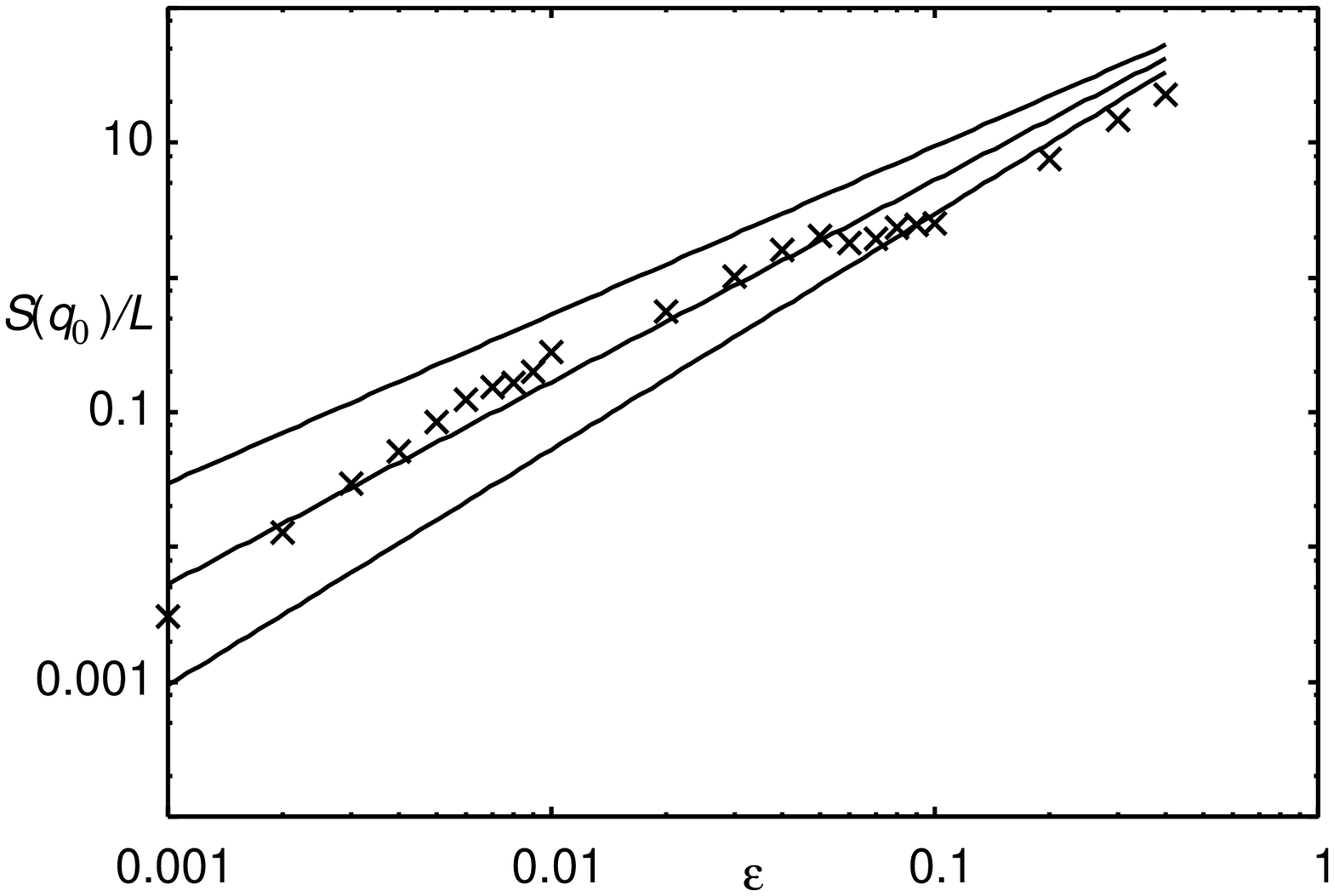}}
\caption{\label{fig:sq0}%
$\epsilon$ dependence of $S(q_0 \simeq 0)$. 
The three lines, included for reference, have slopes of $5/4, 6/4$, and $7/4$.
}
\end{figure}
\begin{figure}
\resizebox{0.45\textwidth}{!}{\includegraphics{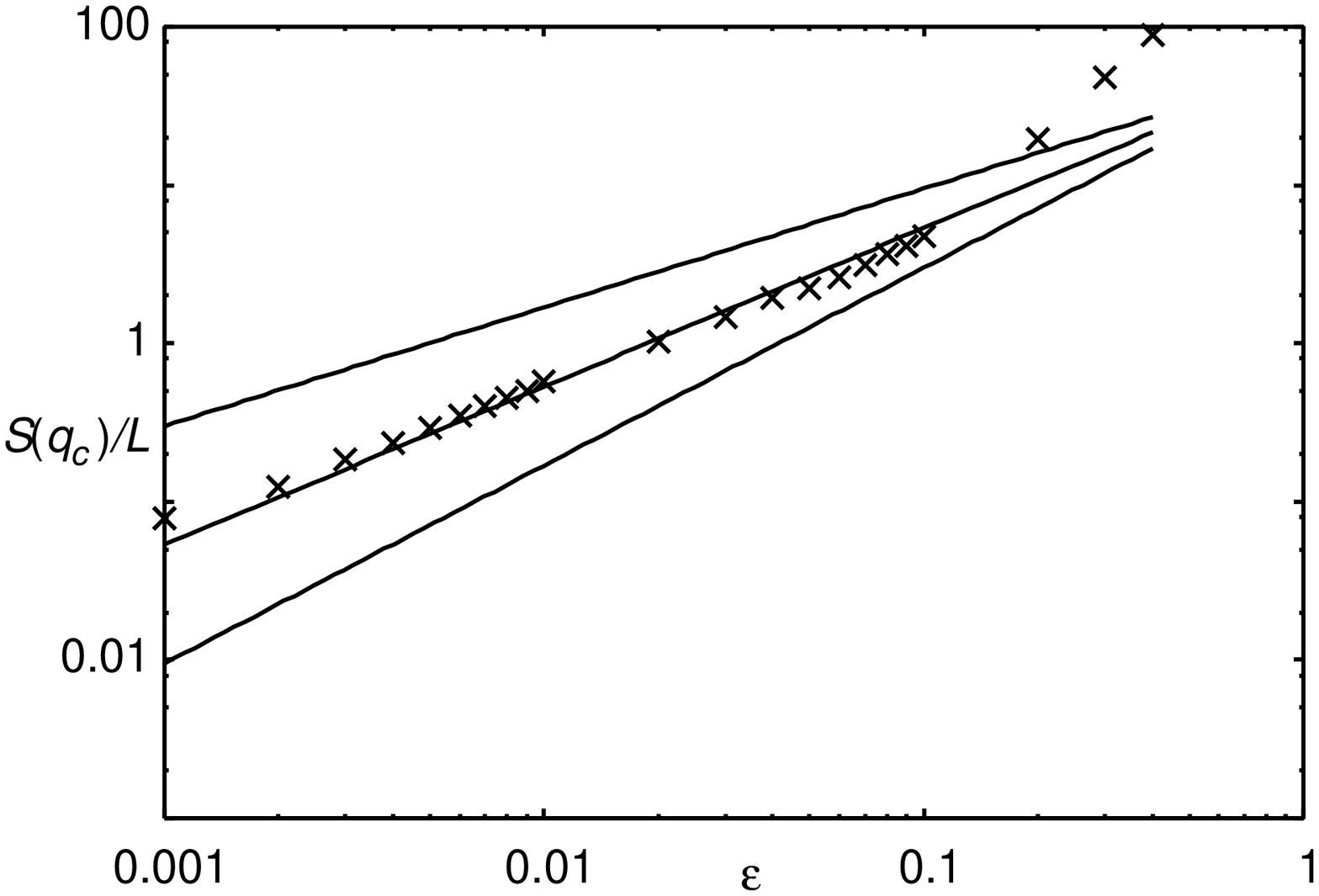}}
\caption{\label{fig:sqc}%
$\epsilon$ dependence of $S(q_c \simeq 1)$. 
The three lines, included for reference, have slopes of $3/4, 4/4$, and $5/4$.
}
\end{figure}
Here, substituting 
$v=\epsilon^\alpha A(X,T) e^{ix}+ c.c. + \epsilon^\beta f(X,T)$ 
into the Wiener-Khintchine relation, 
\begin{equation}
S(q)=L\int^L_0  \langle v(x)v(0) \rangle e^{-iqx}dx, \label{3}
\end{equation}
we obtain 
\begin{equation}
S(1) =  L \int^L_0 \epsilon^{2\alpha}  \langle A(X)\bar{A}(0) \rangle  dx, \label{4} 
\end{equation}
where we have used $ \langle AA \rangle = \langle \bar{A}\bar{A} \rangle = \langle Af \rangle = \langle \bar{A}f \rangle =0$. 
Now, because the characteristic spatial scale of turbulent fluctuations is 
$\epsilon^{-1/2}$, as shown in \F{dq}, we can reasonably assume 
$ \langle A(X)\bar{A}(0) \rangle  = \exp[- \epsilon^{1/2} x]$. 
Then, substituting this into the integrand of \E{4}, we obtain 
\begin{equation}
S(1)=L \epsilon^{2\alpha-1/2} \label{5}. 
\end{equation}
Comparing this equation and the result displayed in \F{sqc}, 
we find $\alpha=3/4$, 
which is consistent with the result found for 
the exponent of $\sqrt{ \langle v^2 \rangle }$ obtained from \F{rms}. 
Similarly, we find
\begin{equation}
S(0)=L \epsilon^{2\beta-1/2} \label{6},  
\end{equation}
with $\beta=1$. 
Thus, we arrive at the following conclusions.
Our results confirm the validity of the form given in \E{2}.
Further, they indicate that the result for the 
exponent of $\sqrt{ \langle v^2 \rangle }$ given in Ref.\cite{Kli}
is erroneous because the value had not yet converged, 
as conjectured above. 
Equations~\ref{5} and \ref{6} imply that  
the amplitudes of the Fourier modes with wavenumbers $2\pi/L$ ($\simeq 0$) and $1$ defined in Ref.\cite{Tri-Tsu} 
are $2\pi \epsilon^{1/4} /\sqrt{L}$ and 
$2\pi \epsilon^{1/2}/\sqrt{L}$, respectively. 
Thus, our results indicate that 
the values of the quantity ${\rm Re} U/\sqrt{\epsilon}$  
in Figs.(3) and (4) in Ref.\cite{Tri-Tsu} 
are of order $1$ and $0.1$, respectively, 
because the parameter values used there are 
$\epsilon=0.0001$ and $L=2\pi/p$, where $p=3.125 \times 10^{-3}$. 
In fact, the figures in Ref.\cite{Tri-Tsu} support this argument. 
We believe that the reason why 
P.~C.~Matthews {\em et al.} reported that the form \E{2} is 
inconsistent with the numerical results in Ref.\cite{Tri-Tsu} 
is that they missed the $-1/2$ appearing in the exponents of 
Eqs.~\ref{5} and \ref{6}, 
which is due to the spatial correlation of turbulent fluctuations. 

In summary, we have found that 
the spatial power spectrum of \E{1} in wavenumber space 
is an extensive quantity. 
The spectrum for $\epsilon \geq O(0.1)$ is 
qualitatively indistinguishable from 
that of the Kuramoto-Sivashinsky (KS) equation. 
We obtained the critical exponents of the turbulent fluctuations for \E{1}, 
and we found that these exponents converge for $\epsilon \sle O(0.1)$. 
Beyond such a value, because the unstable and weakly stable modes of 
this equation are not well separated, it exhibits spatiotemporal chaos 
of the KS type, not the Nikolaevskii type. 
Therefore, we conclude that the works presented 
in Refs.~\cite{Xi-Ext, Tor, Kli}, where values of $\epsilon$ greater than 
$0.1$ were used, in fact studied KS-type chaos.  
The numerical results obtained in this paper are 
consistent with those given in Ref.\cite{Tri-Tsu} and 
with the amplitude equations appearing in Ref.\cite{Mat}. 
S.~Toh reported that a pulse-distributed model reproduces 
the spatial spectrum of the KS equation \cite{Toh}. 
We believe that that model is applicable also to \E{1}, 
for which pulses become more regulary distributed as $\epsilon$ decreases. 
The spectrum of the KS equation possesses 
a wavy structure for large wavenumbers. 
We believe that for the spectrum of \E{1}, the peaks of this wavy 
structure, which appear at $q = 1, 2, 3, \cdots$, 
become increasingly sharp as $\epsilon$ decreases.

D.~T.~ is very grateful to H.~Fujisaka for useful discussions 
and gratefully acknowledges 
financial support by the Japan Society for the Promotion of Science (JSPS).

\end{document}